\documentstyle[a4,11pt]{article}
\begin{document}
\begin{titlepage}
\vskip 2cm
\begin{flushright}
Preprint CNLP-1996-02
\end{flushright}
\vskip 2cm
\begin{center}
{\bf
ON THE  M-XX EQUATION}
\footnote{Preprint
CNLP-1996-02. Alma-Ata. 1996 }
\end{center}
\vskip 2cm
\begin{center}
Ratbay MYRZAKULOV
\footnote{E-mail: cnlpmyra@satsun.sci.kz}
\end{center}
\vskip 1cm

\begin{center}
 Centre for Nonlinear Problems, PO Box 30, 480035, Alma-Ata-35, Kazakhstan
\end{center}

\begin{abstract}
Some exact solutions  of the (2+1)-dimensional
integrable classical continuous isotropic
Heisenberg spin chain (the M-XX equation) are obtained by using
Hirota's method.
These solutions are characterized by an integer topological charge.

\end{abstract}

%\maketitle

\end{titlepage}

\setcounter{page}{1}
\newpage

\tableofcontents

\section{Introduction}
Classical continuum Heisenberg ferromagnet models (CCHFM)
exhibit a
rich variety of nonlinear behaviour. In particular, over the past two
decades, several integrable nonlinear ferromagnetic models  have
been idenfied [1-3].  So, some number integrable spin systems
in 2+1 dimensions are found [4-8]. The
underlying nonlinear spin excitations in such spin systems represented
by solitons, domain
walls, vortecies,
lumps and dromions [4-11]. Theirs study are of considerable intrinsic interest,
especially from the points of view of both mathematics and physics.
Integrable spin systems in 2+1 dimensions as their (1+1)-dimensional
counterparts, display
fascinating geometrical aspects: they are gauge and L-equivalent to the
nonlinear
Schr${\ddot o}$dinger - type equations (the Davey-Stewartson equation,
the Zakharov equations (ZE), the Strachan equation and so on) [5, 9-10, 12].
Generally speaking, between spin systems and the differential geometry  take
places the deep connection [12, 13-17, 20]. An important feature of two dimensional
spin systems is the existence  the  topological invariant
$$
Q = \frac{1}{4\pi} \int\int dxdy {\bf S}\dot ({\bf S}_{x} \wedge {\bf S}_{y}) \eqno (1)
$$
such that the solutions of them are classified by the topological charge (1).
In (9) ${\bf S} = (S_{1}, S_{2}, S_{3})$ is the three-dimensional unit vector
(the spin vector). The class of exact solutions of (2+1)-dimensional spin
systems is a very rich one.  The solitons, vortices, dromions, lumps
are among them. These solutions have the important physical significance.
So, for example, vortecies play an active role in the dynamics and
termodynamics of quasi-two-dimensional magnets [18-20].

The present paper is devoted to the study of the
following  CCHFM in two
space and one time dimensions
$$
{\bf S}_t +  {\bf S}\wedge \{(b+1) {\bf S}_{\xi \xi} -b{\bf S}_{\eta \eta}\} +
bu_{\eta} {\bf S}_{\eta} + (b+1)u_{\xi}{\bf S}_{\xi} = 0 \eqno(2a)
$$
$$
u_{\xi \eta} = {\bf S}({\bf S}_{\xi}\wedge {\bf S}_{\eta}) \eqno(2b)
$$
with
$$
\xi = \frac{x}{2} + \frac{a+1}{\alpha}y, \quad \eta = -\frac{x}{2} -
\frac{a}{\alpha}y.    \eqno(3)
$$
where $a, b $ are real constants, $\alpha^{2} = \pm 1$.
It is  the Myrzakulov XX (M-XX) equation [5].
Hereafter, for convenience,  we use
the conditional notations, e.g. equation (2) we denote by
the M-XX equation.
Equation (2) is  integrable. We will distinguish the two integrable cases:
the M-XXA equation as
$\alpha^{2} = 1$ and the M-XXB equation as $\alpha^{2} = -1$. Also,
equation (2) contains several integrable cases: \\
(i) $b = 0$,  yields the M-VIII equation
$$
{\bf S}_t +  {\bf S}\wedge  {\bf S}_{\xi \xi}  + w{\bf S}_{\xi} = 0 \eqno(4a)
$$
$$
w_{ \eta} = {\bf S}({\bf S}_{\xi}\wedge {\bf S}_{\eta}) \eqno(4b)
$$
where $w=u_{\xi}$.

(ii) $a = b = - \frac{1}{2}$, yields the celebrated Ishimori equation [4]
$$
{\bf S}_t +  \frac{1}{2}{\bf S}\wedge \{ {\bf S}_{\xi \xi} +
{\bf S}_{\eta \eta}\} -\frac{1}{2}u_{\eta} {\bf S}_{\eta} + \frac{1}{2}u_{\xi}{\bf S}_{\xi} = 0 \eqno(5a)
$$
$$
u_{\xi \eta} = {\bf S}({\bf S}_{\xi}\wedge {\bf S}_{\eta}) \eqno(5b)
$$
where $\xi = \frac{1}{2}(x  + \frac{1}{\alpha}y), \quad \eta =
-\frac{1}{2}(x -\frac{1}{\alpha}y).$ The Ishimori equation (5) is the first
integrable spin system in plan, which can be solved by the inverse scattering
method (IST). This equation were studied by the different authors from
variety points of view (see, e.g. [4,6-7,9,11,21]).

(iii) Let now we put $b = 0, \eta =t$,  then equation (2) reduces to
the (1+1)-dimensional M-XXXIV equation
$$
{\bf S}_t +  {\bf S}\wedge  {\bf S}_{\xi \xi}  + w{\bf S}_{\xi} = 0 \eqno(6a)
$$
$$
w_{ t} + \frac{1}{2} ({\bf S}^{2}_{\xi})_{\xi} = 0 \eqno(6b)
$$
This integrable equation was introduced in [5] to describe nonlinear dynamics of
compressible magnets.

Equation (2) is the (2+1)-dimensional
integrable generalization  of the (1+1)-dimensional CCHFM or the isotropic
Landau-Lifshitz equation (LLE)
$$
{\bf S}_t = {\bf S} \wedge {\bf S}_{xx}  \eqno(7)
$$
and in 1+1 dimensions reduces to it.  Here, it should be mentioned that
the M-XX equation (2) is not the only  integrable generalization of
the LLE (7) in 2+1 dimensions. There exist several another integrable
generalizations, e.g the following one,
$$
{\bf S}_t = ( {\bf S} \wedge {\bf S}_{y} + u {\bf S})_x   \eqno(8a)
$$
$$
u_{x} = - {\bf S}({\bf S}_x \wedge {\bf S}_y) \eqno(8b)
$$
This equation, which is known as the Myrzakulov I (M-I) equation,
is again completely integrable [5,10,12].

As integrable, equation (2) can be solved by the IST method.
The applicability of the IST method to the M-XX
equation (2) is based on its equivalence to the compatibility condition
of the following linear equations [5]
$$
\Phi_{Z^{+}} = S\Phi_{Z^{-}} \eqno(9a)
$$
$$
\Phi_{t} = 2i[S+(2b+1)I]\Phi_{Z^{-}Z^{-}} +  W\Phi_{Z^{-}} \eqno(9b)
$$
where  $Z^{\pm} = \xi \pm \eta$ and
$$ W = 2i\{(2b+1)(F^{+} + F^{-} S) +(F^{+}S + F^{-}) +
(2b+1)SS_{Z^{-}}+\frac{1}{2}S_{Z^{-}} + \frac{1}{2} SS_{Z^{+}} \}, \quad
$$
$$
S= \pmatrix{
S_3 & rS^- \cr
rS^+ & -S_3
},\quad S^{\pm}=S_{1}\pm iS_{2} \quad  S^2 = EI,\quad E = \pm 1,
\quad r^{2}=\pm 1,
$$
$$
F^{+} = 2iu_{Z^{-}}, \quad F^{-}=2iu_{Z^{+}}.
$$
In fact, from the condition $\Phi_{Z^{+}t} = \Phi_{tZ^{+}}$, we get the equation
$$
iS_t +  \frac{1}{2}[S,(b+1) S_{\xi \xi} -bS_{\eta \eta}] +
ibu_{\eta} S_{\eta} + i(b+1)u_{\xi}S_{\xi} = 0 \eqno(10a)
$$
$$
u_{\xi \eta} =  \frac{1}{4i}tr(S[S_{\xi},S_{\eta}]) \eqno(10b)
$$
which is the matrix form of equation (2).

\section{Bilinearization}

It could be of interest to study the equation (2) by the IST method. But
to look for the some special solutions,
it is convenient use the other may be more practical method - the Hirota
bilinear method. Remaining the use of the IST method in future, in this
paper, we work with the Hirota method. To this purpose, we construct the
bilinear form of (2) as $E=r=1$. Let us we now introduce the following
transformation for the components of spin vector ${\bf S}$
and for the derivatives of scalar potential $u$
$$
S^{+} = \frac{2\bar f g}{\bar f f + \bar g g}, \quad
S_3 = \frac{\bar f f - \bar g g}{\bar f f + \bar g g}, \eqno (11a)
$$
$$
u_{\xi} =- 2i\frac{D_{\xi}(\bar f\circ f +
\bar g \circ g)}{\bar f f +
\bar g g}, \quad
u_{\eta}=  2i  \frac{D_{\eta}(\bar f \circ f +
\bar g \circ g)}{\bar f f + \bar g g}  \eqno (11b)
$$
Here the Hirota operators $D_{x}, D_{y}$ and $D_{t}$ are defined by
$$
D^{l}_{\xi}D^{m}_{\eta}D^{n}_{t} f(\xi, \eta, t)\circ g(\xi, \eta, t) =
(\partial_{\xi}-\partial_{\xi^{\prime}})^{l}
(\partial_{\eta}-\partial_{\eta^{\prime}})^{m}
(\partial_{t}-\partial_{t^{\prime}})^{n} f(\xi, \eta, t)\circ g(\xi^{\prime},
\eta^{\prime}, t^{\prime})\mid_{\xi=\xi^{\prime}, \eta = \eta^{\prime},
t=t^{\prime}} .  \eqno(12)
$$

Substituting the formulae (11) into the M-XX equation (2), we obtain
the bilinear equations
$$
[iD_{t} - (b+1)D_{\xi}^{2} +bD_{\eta}^{2}](\bar f \circ g) = 0  \eqno (13a)
$$
$$
[iD_{t} - (b+1)D_{\xi}^{2} +bD_{\eta}^{2}](\bar f \circ f - \bar g \circ g) = 0  \eqno (13b)
$$
$$
\{D_{\xi} D_{\eta}+D_{\eta} D_{\xi}\} (\bar f  f + \bar g  g) \circ
(\bar f f + \bar g  g) = 0 \eqno (13c)
$$
Note that equation (13c) coincide with  the compatibility condition $ u_{\xi\eta} = u_{\eta\xi} $.
Equations (13) is the desired Hirota bilinear form of equation (2).

\section{Solutions}
Now we can construct some special solutions of equation (2). As examples,
we find simplest soliton, domain wall and vortex solutions.

\subsection{Soliton  solution}
\begin{center}
{\bf FIND THE SOLITON SOLUTIONS.}
\end{center}

\subsection{Domain wall  solution}
In order to obtain a domain wall solutions,
we make the choice
$$
f=1. \eqno (14)
$$
Then, equations (13a,b) reduce to
$$
ig_{t}+(b+1)g_{\xi\xi} - b g_{\eta\eta} = 0  \eqno(15a)
$$
$$
(b+1)\bar g_{\xi}g_{\xi} - b\bar g_{\eta}g_{\eta} = 0 \eqno(15b)
$$
Let us we consider the case, when $\alpha^{2} = -1$, i.e. the M-XXB equation.
In this case,  equation (13c) is identically satisfied by any analytical
function  $g = g(\xi, \eta, t)$. For example, the simplest non-trivial solution
of equation (2) is
$$
g = \exp \chi_{1}  \eqno (16)
$$
where
$$
\chi_{1} = m_{1}\xi + n_{1}\eta + i [(b+1)m^{2}-bn^{2}]t + \chi_{10} =
\chi_{1R}+i\chi_{1I}  \eqno (17)
$$
Thus, the spin components and the potential field are given by
$$
S^{+} = e^{i\chi_{1I}} sech \chi_{1R},\quad
S_{3} = - th\chi_{1R},\quad  u = 2\ln (1+e^{2\chi_{1R}}). \eqno (18)
$$

\subsection{ Vortex  solution}.
To construct vortex solution, we use the equation (13) and assume that its
solution has the form
$$
f=f(\xi, t), \quad g = g(\xi, t)  \eqno(19)
$$
Then the condition (13c) is satisfied automatically. At the same time,
equations (13a,b) are satisfy if
$$
if_{t}+ (b+1) f_{\xi\xi} = 0  \quad ig_{t} + (b+1) g_{\xi \xi} = 0 \eqno(20)
$$
Hence, we obtain the following multi-vortex  solutions of the M-XXB equation
(2)
$$
g_{N} = \sum_{j=0}^{N}\sum_{m+2n=j}\frac{a_{j}}{m!n!}(\frac{2}{b+1})^{\frac{m}{2}}\xi^{m}(2it)^{n}  \eqno(21a)
$$
$$
f_{N} = \sum_{j=0}^{N-1}\sum_{m+2n=j}\frac{b_{j}}{m!n!}(\frac{2}{b+1})^{\frac{m}{2}}\xi^{m}(2it)^{n}  \eqno(21b)
$$
where $a_{j}$ and $b_{j}$ are arbitrary complex constants, $m,n$ are the
non-negative integer numbers. In particular, the 1-vortex solution isgiven by
$$
f=b_{0}, \quad g=a^{\prime}_{1}\xi + a_{0} \eqno (22)
$$
where $a^{\prime}_{1}=a_{1}(\frac{2}{b+1})^{\frac{1}{2}}$. The corresponding solution
of equation (2) is given by
$$
S^{+} = \frac{2b_{0}(a^{\prime}_{1}\xi + a_{0})}{\mid b_{0}\mid^{2} +
\mid a^{\prime}_{1}\xi + a_{0}\mid^{2}}
 \eqno (23a)
$$
$$
S_{3} =  \frac{
\mid b_{0}\mid^{2} - \mid a^{\prime}_{1}\xi + a_{0}\mid^{2}}
{\mid b_{0}\mid^{2} + \mid a^{\prime}_{1}\xi + a_{0}\mid^{2}}
\eqno (23b)
$$
$$
u =  2\ln (\mid b_{0}\mid^{2} + \mid a^{\prime}_{1}\xi + a_{0}\mid^{2})
\eqno (23c)
$$
So, the 1-vortex solution is static. To find the dynamic solution,
we consider the 2-vortex solution, which  has the form
$$
f=b^{\prime}_{0}\xi + b_{0}, \quad g=\frac{a_{2}}{b+1}\xi^{2} +
\frac{a_{2}}{2} 2it+
a^{\prime}_{1}\xi + a_{0}, \quad b^{\prime}_{1}=b_{1}(\frac{2}{b+1})^{\frac{1}{2}}
\eqno (24)
$$

The interesting question is the dynamics of these vortices. Let
us rewrite the solution (21) in the following factorized form
$$
f(\xi, t) = b_{0} \prod^{N}_{j=1}[\xi - p_{j}(t)]  \eqno(25a)
$$
$$
g(\xi, t) = a_{0} \prod^{N}_{j=1}[\xi - q_{j}(t)]  \eqno(25b)
$$
where $p_{j}$ and $q_{j}$ denote the positions of the zeros for $f$ and $g$,
and $a_{0}, b_{0}$ are constants. Substituting (25) into (20), we get the
evolutions of $p_{j}$ and $q_{j}$ as
$$
p_{jt} = -i(b+1)\sum^{N}_{k\not= j}\frac{1}{p_{j}-p_{k}}   \eqno(26a)
$$
$$
q_{jt} = -i(b+1)\sum^{N}_{k\not= j}\frac{1}{q_{j}-q_{k}}   \eqno(26b)
$$
where $j,k=1,2,...,N$. Hence, we get the Calogero-Moser type system
$$
p_{jtt} = 2(b+1)^{2}\sum^{N}_{k\not= j}\frac{1}{(p_{j}-p_{k})^{3}}   \eqno(27a)
$$
$$
q_{jtt} = 2(b+1)^{2}\sum^{N}_{k\not= j}\frac{1}{(q_{j}-q_{k})^{3}}   \eqno(27b)
$$
with the following Hamiltonian
$$
H=\frac{1}{2} \sum (p_{jt}^{2}+q^{2}_{jt}) + (b+1)^{2}\sum [ (p_{j}-p_{k})^{-2}
+(q_{j}-q_{k})^{-2}].    \eqno(28)
$$

\subsection{ Dromion   solution}
In this subsection we would like get the dromion [23] solution of equation (2),
but please
\begin{center}
{\bf FIND THE DROMION SOLUTIONS.}
\end{center}

\section{Gauge equivalent equation}

Finally, let us we present the gauge equivalent counterpart of equation (2).
It has the form
$$
iq_t+ (1 + b)q_{\xi \xi } - b q_{\eta \eta } + vq = 0 \eqno(29a)
$$
$$
ip_t - (1 + b)p_{\xi \xi } + b p_{\eta \eta } - vp = 0 \eqno(29b)
$$
$$
v_{\xi \eta } = -2\{(1+ b) (pq)_{\xi \xi} - b(pq)_{\eta \eta}\} \eqno(29c)
$$
where $p,q$ are some complex functions.  Equation (29) is related with the
Zakharov equations [22].  To prove gauge equivalence between equations
(2) and (29), let us perform the
gauge transformation $\Psi = g \Phi$, where the function $\Phi$ is
the solution of equations (9) and $g$ is a 2x2 matrix such that
$$
S = g^{-1}\sigma_{3}g   \eqno (30)
$$
and
$$
g_{Z^{+}}g^{-1} - \sigma_{3} g_{Z^{-}}g^{-1}
=
\left ( \begin{array}{cc}
0   & q \\
p   & 0
\end{array} \right)   \eqno(31)
$$
Under this transformation the function $\Psi$ satisfies the following
set of linear equations
$$
\Psi_{Z^{+}} = \sigma_{3} \Psi_{Z^{-}} + B_{0}\Psi \eqno(32a)
$$
$$
\Psi_{t} = 4i C_{2} \Psi_{Z^{-}Z^{-}} + 2 C_{1} \Psi_{Z^{-}} + C_{0}\Psi. \eqno(32b)
$$
where $B_{0},\quad C_{j}$ are given by
$$
B_{0}= \pmatrix{
0   &  q \cr
p   &  0
}, \quad
C_{2}= \pmatrix{
b+1 & 0 \cr
0   & b
},\quad
C_{1}= \pmatrix{
0   &  iq \cr
ip  &  0
},\quad
C_{0}= \pmatrix{
c_{11}  &  c_{12} \cr
c_{21}  &  c_{22}
}
$$
$$
c_{12}=i[(4b+3)q_{Z^{-}}+q_{Z^{+}}] \quad
c_{21}=-i[(4b+1)p_{Z^{-}}+ p_{Z^{+}}]
$$
and $v=i(c_{22}-c_{11})$. Here $c_{jj}$ are the solution of the  following
equations
$$
c_{11Z^{-}}- c_{11Z^{+}} = i[(4b+3)(pq)_{Z^{-}} +  (pq)_{Z^{+}}]
$$
$$
c_{22Z^{-}}+ c_{22Z^{+}} = i[(4b+1)(pq)_{Z^{-}} + (pq)_{Z^{+}}].
$$

The compatibility condition of equations (32) gives the equation (29). Therefore
the M-XX equation (2) and the equation (29) are gauge equivalent to each other.
Now let us proceed to the M-XX equation (10). It is not difficult to
check that if $g$ obeys equations (31) then the $S$ in the form (30) obeys
the M-XX equation (10) with
$$
u = -2i\alpha \ln \det g   \eqno(33)
$$

\section{Conclusion}

To conclude, in this paper we have found some exact solutions, namely
domain wall and vortex solutions of the (2+1)-dimensional CCHFM - the
M-XX equation. We have shown that the dynamics of vortices are governed by the
Calogero-Moser type systems. Also we have presented the gauge equivalent counterpart of this
equation.

\section{Particular open problems}

Finally, we also would like to pose the following particular problems: \\
{\bf Problem N1:}  Find solutions of the M-XX equation by
the $\bar \partial$-dressing method. \\
{\bf Problem N2:}  Find solutions of the M-XX equation using
the nonlocal Riemann-Hilbert problems method. \\
{\bf Problem N3:}  Find solutions of the M-XX equation by
means of the Darboux transformation as in [21]. \\
{\bf Problem N4:}  Find the other solutions of the M-XX equation
(solitons, dromions and so on) by the Hirota bilinear method.\\

If you have some results in these directions, please, inform me by
E-mail: cnlpmyra@satsun.sci.kz. We are ready to interaction.

\section{Acknowledgments}

The author  would like to thank  Prof. M.Lakshmanan for hospitality
during his visit to  Bharathidasan University and for stimulating
discussions.
He is grateful for helpful conversations with
A.Kundu, Radha Balakrishnan, M.Daniel, R.Radha and  R.Amuda.
Also he  would like to thank Dr.
Radha Balakrishnan for hospitality during his visit to the Institute of
Mathematical Sciences.

\end{document}